\documentclass[12pt,eadjoint tfnpsf]{article}

\catcode`\@=11
\@addtoreset{equation}{section}

\usepackage{amsbsy,amssymb,latexsym,amsfonts}
\usepackage{graphicx}
\usepackage{amsbsy,amssymb,latexsym,amsfonts}
\usepackage{amsmath,epsfig}
\usepackage{bm}
\usepackage{graphicx}

\RequirePackage[dvips,usenames]{color}
\definecolor{fireblick}{rgb}{0.698039,0.133333,0.133333}

\newcommand{\beq}{\begin{equation}}
\newcommand{\eeq}{\end{equation}}
\newcommand{\bea}{\begin{eqnarray}}
\newcommand{\eea}{\end{eqnarray}}

\newcommand{\CN}{{\mathcal N}}

\newcommand{\CS}{{\mathcal S}}

\newcommand\ad{{\rm ad}}

\newcommand\tr{\mathrm{tr}}

\numberwithin{equation}{section}
\newcommand{\bel}[1]{\begin{equation}\label{#1}}                     
\newcommand{\bal}[1]{\begin{eqnarray}\label{#1}}                     
\newcommand{\be}{\begin{equation}}
\newcommand{\ee}{\end{equation}}
\def\beq{\begin{equation}}
\def\eeq{\end{equation}}

\newcommand{\mat}[1]{\begin{pmatrix} #1 \end{pmatrix}}
\renewcommand{\thefootnote}{\fnsymbol{footnote}}

\newcommand{\ol}[1]{\overline{#1}}

\def\a{\alpha}

\def\e{\epsilon}
 
\def\l{\lambda}
\def\s{\sigma}
\def\z{\zeta}

\def\G{\Gamma}

\def\Pst{\tilde{\Psi}}

\def\CS{{\mathcal S}}

\def\CN{\mathcal N}
\def\CT{\mathcal T}

\def\12{\frac{1}{2}}

\def\eff{\mathrm{eff}}

\def\tr{\mathrm{tr}}

\def\ad{\text{ad}}
\def\as{\text{as}}

\def\vt{\tilde{v}}
\def\xt{\tilde{x}}
\def\xit{\tilde{\xi}}

\def\CTt{\tilde{\mathcal{T}}}
\def\CSt{\tilde{\mathcal{S}}}

\begin{document}
%
%
\begin{titlepage}

\begin{flushright}
\normalsize
~~~~
April, 2009 \\
OCU-PHYS 313 \\
\end{flushright}

\bigskip
\bigskip

\begin{center}
{\Large\bf   Orientifolded Matrices and Supersymmetries that Give Rise to 
Spacetime Directional Asymmetry of Effective Interactions}
\end{center}

\bigskip

\begin{center}
{%
H. Itoyama$^{a,b}$\footnote{e-mail: itoyama@sci.osaka-cu.ac.jp}
and
R. Yoshioka$^b$\footnote{e-mail: yoshioka@sci.osaka-cu.ac.jp}
}
\end{center}

\bigskip

\begin{center}
$^a$ \it Department of Mathematics and Physics,
Graduate School of Science\\
Osaka City University\\

\medskip

$^b$ \it Osaka City University Advanced Mathematical Institute (OCAMI)

\bigskip

3-3-138, Sugimoto, Sumiyoshi, Osaka, 558-8585, Japan \\

\end{center}


\bigskip

\begin{abstract}
Effects of matrix orientifolding that preserves supersymmetries 
 are considered in the IIB matrix model with regard to its effective dynamics 
 generated for diagonal elements. 
 Taking the case of maximal supersymmetries and  the long distance expansion 
 of the one-loop effective action as well as cases where the size of the matrices is small,
 we demonstrate that the directional asymmetry of spacetime brought upon by this setup 
  in fact leads to that of the forces exerting on the spacetime points:
   in addition to the two-body attraction between two points, there are attractions toward 
   the four dimensional plate.
\end{abstract}

\vfill

\setcounter{footnote}{0}
\renewcommand{\thefootnote}{\arabic{footnote}}

\end{titlepage}

\section{Introduction}

Continuing attention has been paid for the reduced matrix models \cite{BFSS}-\cite{KSlv} 
 where one intends to 
 carry out spacetime formation and their interplay with particle physics.
They are obtained from the ten dimensional Yang-Mills  theory and it cousins by 
 the dimensional reduction to  zero, one and two dimensions. 
 (For a review, for example, \cite{Tay}.)
Several operations are known to be materialized on matrices without approximation. 
In particular, the twist operation which renders the first quantized oriented
 theory into nonorientable one  has  a precise matrix counterpart. 
Including this matrix twist upon $Z_2$ identification, we obtain matrix construction 
 of $Z_2$ orientifold (in the large radius limit) as well \cite{IToku}. 
What all these mean is  simply that a matrix representing an element of 
 $u(2k)$ Lie algebra becomes a sum of the matrix in the antisymmetric 
 representation of $usp(2k)$ and that representing an element of $usp(2k)$ 
 Lie algebra \cite{IToku}-\cite{DF}. 
The Chan-Paton factor of orthogonal group is then generated as 
 gauge symmetry \cite{ITsu}. 
      
On the other hand,  it has been found to be quite nontrivial to maintain supersymmetry
 upon this operation at the matrix level \cite{IToku,IY}. 
For example, taking all bosonic matrices lying 
 in the antisymmetric representation, 
 which would represent ten-dimensionally covariant unorientied theory, is simply not 
 allowed by a property of representation: $[antisym, antisym] \sim adj$. 
Taking all in the defining representation of the $usp(2k)$ Lie algebra 
 has no way to satisfy the second supersymmetries called kinematical supersymmetries, 
 which simply shift  any fermionic matrix by an amount proportional to the unit matrix. 
The upshot is that matrix orientifolding inevitably introduces spacetime directional 
 asymmetry.
Orbifolding \cite{AIS} and orientifolding of matrices have also played roles 
 in the recent studies \cite{Un} of lattice supersymmetry 
 and its large N Eguchi-Kawai reduction \cite{EK}. 
    
A central theme is  the effective dynamics for the diagonal elements of the matrices
 in particular, that for the bosonic ones, which are obtained by the integrations 
 of the off-diagonal elements.\footnote{For exact integrations of partition functions, 
 see \cite{MNS}-\cite{Kih}.}
The forces exerting among them  are supposed to dictate  formation of our spacetime
 consisting of points. 
The goal of this paper is to see the effect of the spacetime directional asymmetry 
 qualitatively and semi-quantitatively at the level of the forces exerting among 
 spacetime points. 
Taking the case of maximal supersymmetry
 which generates $SO(2n)$ Chan-Paton factor, namely, the case of 
 the $USp$ matrix model \cite{IToku},
 we will show  that the points not only attract each other 
 by the two-body force \cite{AIKKT}  shared by the IIB matrix model before orientifolding, 
 but also they get attracted toward the four dimensional plate.
Thus this offers a simple mechanism for generating our four dimensional spacetime 
 from matrices and agrees in conclusion with that drawn sometime ago in \cite{CIK2} 
 based on the Yang-monopole and the higher dimensional analog derived 
 from the nonabelian Berry phase \cite{IM}. (See also \cite{GT,PST}.)
 
In the next section, we briefly review the matrix orientifolding 
 with maximal supersymmetries in the IIB matrix model.  
The branching of the $u(2k)$ Lie algebra  into the $usp(2k)$ Lie algebra and 
 the antisymmetric representation is recalled.
In section three, we present  the long distance expansion of the one-loop effective
 action of the diagonal elements. 
While there are several robust properties of the effective action noted in \cite{AIKKT} 
 which are preserved by the matrix orientifolding, 
 we will focus upon  new features brought upon by this operation. 
In section four, we consider a few cases in which the size of the matrices is small, 
 check the consistency with section three and draw the conclusion stated above.

\section{IIB matrix model and matrix orientifolding}

Recall the IIB matrix model  \cite{IKKT} 
 which is obtained by the dimensional reduction 
 of ten dimensional $\CN = 1$ supersymmetric Yang-Mills theory to a point: 
\begin{align}
 S  &= \tr 
 \left( - \frac{1}{4g^2} [v_M , v_N]^2 
 -\frac{1}{2g^2} \bar{\Psi} \Gamma^{M} [ {v_{M}} , \Psi ]   \right), \cr
 & = S_b + S_f, 
\label{action} \\
S_b &= - \frac{1}{4g^2} \tr [v_M , v_N]^2, \\
S_f &= - \frac{1}{2g^2} \tr \bar{\Psi} \Gamma^{M} [ {v_{M}} , \Psi ].
\end{align}
The fermionic matrix $\Psi$ is a ten dimensional Majorana-Weyl spinor. 
The action (\ref{action}) has 16 + 16 supersymmetries: 
\begin{align}
\delta_{\text{IIB}}^{(1)}{\Psi}
&=\frac{i}{2}[{v}_M,{v}_N]\Gamma^{MN} \e \\
\delta_{\text{IIB}}^{(1)}{v}_M 
&=i\bar{\epsilon}
\Gamma^M {\Psi} \\
\delta^{(2)}_{\text{IIB}}{\Psi}
& =\xi \\
\delta^{(2)}_{\text{IIB}}{v}_M 
& =0
\end{align}
where $\e$ and $\xi$ are the grassmann parameters of these transformations.

Orientifolding at the level of matrices was given in ref.{\cite{IToku}}, 
 where the branching of  $u(2k)$ Lie algebra into $usp(2k)$ Lie algebra 
 and the antisymmetric representation of $usp(2k)$ is exploited. 
Let us recall this more explicitly. The $u(2k)$ Lie algebra splits 
 into two representations of the $usp(2k)$ Lie algebra: 
\begin{align}
{\mathfrak{adj}}(2k)=\{ X\in\mathfrak{u}(2k)|X^tF+FX=0\}, \\ 
{\mathfrak{asym}}(2k)=\{ X\in\mathfrak{u}(2k)|X^tF-FX=0\}. 
\end{align}
Here  
\begin{equation}
F= 
\begin{pmatrix}
0&I_k\\
-I_k&0
\end{pmatrix},
\end{equation} 
with $I_k$ is $k\times k$ unit matrix.

It is expedient to introduce the following projection and act either  $\hat{\rho}_{-}$ or 
 $\hat{\rho}_{+}$ on each matrix \cite{IToku}: 
\begin{equation}
\hat{\rho}_{\mp}\bullet=\frac{1}{2}(\bullet\mp F^{-1}\bullet^tF). 
\label{projector}
\end{equation}
Requiring to preserve $8+8$ supersymmetries, 
  we obtain \cite{IToku}
\bea
  \{ v_M \} = \{ v_{\mu}, v_{m} \},  \mu = 0,1,2,3,4,7, \;\; m = 5,6,8,9 \nonumber \\ 
  v_{\mu} \in\mathfrak{adj}(2k), \;\;  v_{m} \in\mathfrak{asym}(2k), 
\eea
  while,  by an appropriate choice of the gamma matrices,
   $\Psi$ is determined  as 
\be
\Psi=(\lambda,0,\psi_{(1)},0,\psi_{(2)},0,\psi_{(3)},0, 
 0,\bar{\lambda},0,\bar{\psi}_{(1)},0,\bar{\psi}_{(2)},0,\bar{\psi}_{(3)})^t, 
\ee
where 
\begin{align}
&\lambda, 
\psi_{(1)},
\bar{\lambda},
\bar{\psi}_{(1)}
\in\mathfrak{adj}(2k), \\
&\psi_{(2)}, 
\psi_{(3)},
\bar{\psi}_{(2)},
\bar{\psi}_{(3)}
\in\mathfrak{asym}(2k). 
\end{align}
Note that splitting of ten dimensions into six and four has taken place already 
 at this level.

\section{Effect of matrix orientifolding at long distance expansion of the
one-loop effective action}

In this section, we consider the one-loop effective action of the diagonal elements 
 for the  IIB matrix model (\ref{action})  which is now orientifolded and 
 maintain maximal supersymmetry. 
Let us first decompose the matrices $v_M$ and $\Psi$ into the diagonal 
 and the off-diagonal parts: 
\be
 v_M = x_M + \vt_M, ~~~~~
 \Psi = \eta + \Pst. 
\ee
The off-diagonal parts $\vt_M$ and $\Pst$ are regarded as  fluctuations. 
For our purposes, 
it is enough to consider the fluctuations up to the quadratic order. 
The bosonic part and the fermionic part are respectively written as  
\begin{align}
 S_b^{(2)} &= - \frac{1}{2g^2} \tr [x_M , \vt_N] [x^M , \vt^N] 
              + \frac{1}{2g^2} \tr [x_M , \vt_N] [x^N , \vt^M], 
 \label{b:2}\\
 S_f^{(2)} &= - \frac{1}{2g^2} \tr \bar{\Pst} \G^M [x_M , \Pst] 
 	      - \frac{1}{2g^2} \tr \bar{\eta} \G^M [\vt_M , \Pst]
              - \frac{1}{2g^2} \tr \bar{\Pst} \G^M [\vt_M , \eta] \cr
 	   &= - \frac{1}{2g^2} \tr \bar{\Pst} \G^M [x_M , \Pst] 
 	      - \frac{1}{2g^2} \tr [\bar{\eta} , \vt_M] \G^M \Pst 
              - \frac{1}{2g^2} \tr \bar{\Pst} \G^M [\vt_M , \eta], 
 \label{f:2}
\end{align}
 where we have used the fact that  $\tr A[B,C] = \tr [A,B] C$ is satisfied 
 when $B$ is a bosonic matrix.

We fix the the $U(2k)$ symmetry by the background gauge. 
The gauge fixing term added to the action is  
\be
 S_{g.f.} = - \frac{1}{2g^2} \tr [x_M , v^M]^2 
          = - \frac{1}{2g^2} \tr [x_M , \vt^M]^2 \equiv S_{g.f.}^{(2)}, 
 \label{gauge fix}
\ee
and the corresponding ghost term is 
\be
 S_{ghost} = -\frac{1}{g^2} \tr [x_M,b][x_M,c],  
\ee
where $c$ and $b$ are the ghost and the anti-ghost respectively.

Using 
\begin{align} 
 \tr [ x_M , \vt^M ] [ x_N , \vt^N] - \tr [x_M , \vt_N] [x^N , \vt^M] 
  &= \tr \{ 2 \vt^M \vt^N [x_M , x_N] \} \cr
  &= 0 ~~~ ( \text{when} ~ [x_M , x_N] = 0 ), 
\end{align}
we  obtain  
\be
 S_b^{(2)} + S_{g.f.} = - \frac{1}{2g^2} \tr [x_M , \vt_N] [x^M , \vt^N].  
\ee
In the set up of section two, the matrices take either of the following two forms: 
 six of the matrices belonging to the defining representation of the $usp(2k)$ Lie algebra 
 are given by
\be
x_{\mu} =  \mat{X_{\mu}  & 0 \\ 0 & -X_{\mu}}, ~~~~
\vt_{\mu} = \mat{M_{\mu} & N_{\mu} \\ N^*_{\mu} & -M^t_{\mu}} 
\ee
with $M^{\dag}_{\mu} = M_{\mu}$ and $N^t_{\mu} = N_{\mu}$ while 
four of those belonging to the antisymmetric representation are given by 
\be
x_{n} =  \mat{X_{n} & 0 \\ 0 & X_{n}}, ~~~~
\vt_{n} = \mat{A_{n} & B_{n} \\ B^{\dag}_{n} & A^t_{n}},
\ee
with $A^{\dag}=A$ and  $B^t=-B$. 
Here, 
\be
X_M = \mat{x_{M}^1 &&& \\ & x_{M}^2 && \\ &&\ddots & \\ &&& x_{M}^k }. 
\ee
Note that the matrices $A_n$ and $M_{\mu}$ are matrices consisting of off-diagonal elements. 
  By construction,  the ten-dimensional Lorentz covariance is broken explicitly. 
  In order to refer easily to which representation each matrix belongs to,  we use
 the different space-time indices.
 Greek letters $\mu$, $\nu$, $\cdots$ are used to label the matrices  belonging to 
  the defining representation of the $usp(2k)$ Lie algebra 
 while Roman letters $n$, $m$, $\cdots$ are used to label the matrices belonging to 
  the antisymmetric representation.
 
\begin{figure}[h]
\begin{center}
\includegraphics[scale=0.6]{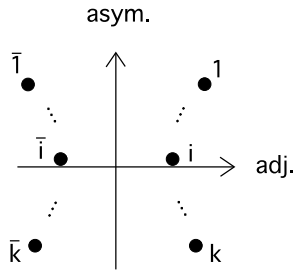}
\caption{
 Each node represents the diagonal elements of the matrices $x_M$, 
 which are regarded as spacetime points and their images in ten dimensional spacetime.
 Because  ten $2k \times 2k$  bosonic matrices appear in the USp matrix model, 
 there are 2$k$ points in this figure.  
 Clearly, the spacetime points and their images are symmetric with respect to the 
 four-dimensional plane spanned by the antisymmetric directions. 
}
\label{fig:diag}
\end{center}
\end{figure}

The upper half part of the diagonal elements of the ten bosonic matrices 
 represents the spacetime points. 
We represent the diagonal matrices $x_M$ as shown in Figure \ref{fig:diag}.  

Exploiting  the relation, 
\begin{align}
 &[X_{M} , A]^{ij} = x_{M}^{ij} A^{ij} \cr
 &\{X_{M} , A\}^{ij} = \xt_{M}^{ij} A^{ij}, 
\end{align}
where $x_M^{ij} = (x_M^i - x_M^j)$ and $\xt_M^{ij} = (x_M^i + x_M^j)$, 
 we obtain 
\begin{align}
[x_{\mu} , \vt_{\nu}] 
 &= \mat{ [X_{\mu} , M_{\nu}] & \{X_{\mu} , N_{\nu}\} \\ 
          -\{X_{\mu} , N^*_{\nu}\} & [X_{\mu} , M^t_{\nu}] } 
  = \mat{ x_{\mu}^{ij} M_{\nu}^{ij} & \xt_{\mu}^{ij} N_{\nu}^{ij}\\
          - \xt_{\mu}^{ij} N_{\nu}^{*ij} & x_{\mu}^{ij} M_{\nu}^{*ij} } 
          ~~~ (\text{adj})
  \label{[ad,ad]} \\
[x_{n} , \vt_{m}] 
 &= \mat{ [X_n , A_m] & [X_n , B_m] \\ 
          [X_n , B^{\dag}_m] & [X_n , A^t_m] }  
  = \mat{ x_n^{ij} A_m^{ij} & x_n^{ij} B_m^{ij} \\ 
          - x_n^{ij} B_m^{*ij} & x_n^{ij} A_m^{*ij}} ~~~ (\text{adj}) 
  \label{[as,as]}\\
[x_{\mu} , \vt_{n}] 
 &= \mat{ [X_{\mu} , A_{n}] & \{X_{\mu} , B_{n}\} \\ 
          -\{X_{\mu} , B^{\dag}_{n}\} & - [X_{\mu} , A^t_{n}] } 
  = \mat{ x_{\mu}^{ij} A_n^{ij} & \xt_{\mu}^{ij} B_n^{ij} \\
  	  \xt_{\mu}^{ij} B_n^{*ij} & - x_{\mu}^{ij} A_n^{*ij} } 
          ~~~ (\text{asym}) 
  \label{[ad,as]} \\
[x_{n} , \vt_{\mu}] 
 &= \mat{ [X_n , M_{\mu}] & [X_n , N_{\mu}] \\ 
          [X_n , N^{*}_{\mu}] & - [X_n , M^t_{\mu}] } 
  = \mat{ x_{n}^{ij} M_{\mu}^{ij} & x_{n}^{ij} N_{\mu}^{ij} \\
  	  x_{n}^{ij} N_{\mu}^{*ij} & - x_{n}^{ij} M_{\mu}^{*ij} } 
          ~~~ (\text{asym}), 
  \label{[as,ad]}
\end{align}
where in the last expression of (\ref{[ad,ad]}) $\sim$ (\ref{[as,ad]}) we have 
displayed each block by its $(i,j)$ element.
After some calculation, we obtain for the bosonic part 
\begin{align}
 S_b^{(2)} + S_{g.f}
   &= \frac{1}{g^2} \sum_{i,j} \left[
   	\{ (x_{\nu}^{ij})^2 + (x_{m}^{ij})^2 \} M_{\mu}^{ij} M^{\mu ij*} 
           + \{ (\xt_{\nu}^{ij})^2 + (x_{m}^{ij})^2 \} N_{\mu}^{ij} N^{\mu ij*} 
        \right. \cr 
        &\hspace{2.3cm} \left.
           + \{ (x_{\nu}^{ij})^2 + (x_{m}^{ij})^2 \} A_{n}^{ij} A^{n ij*}
           + \{ (\xt_{\nu}^{ij})^2 + (x_{m}^{ij})^2 \} B_{n}^{ij} B^{n ij*}
        \right]. 
\end{align}
Note that, for $M_{\mu}^{ij}M^{\mu ij*}$ terms and $A_{m}^{ij}A^{m ij*}$ terms, 
 $i=j$ contributions automatically drop.
We proceed to the fermionic part in a similar fashion. 
Let us write the matrix belonging to the defining representation 
 of the $usp(2k)$ Lie algebra as 
\begin{eqnarray}
 \Pst_{\a} = \mat{O_{\a} & P_{\a} \\ P^*_{\a} & -O_{\a}^t}, &&
 \eta_{\a} = \mat{\Xi_{\a} & 0 \\ 0 & -\Xi_{\a} }, 
 \label{spinor:ad}
\end{eqnarray}
with $O^{\dag}=O$ and $P^t=P$ and that belonging to the 
 antisymmetric representation as
\begin{eqnarray}
 \Pst_{\a'} = \mat{C_{\a'} & D_{\a'} \\ D^{\dag}_{\a'} & C_{\a'}^t}, &&
 \eta_{\a'} = \mat{Z_{\a'} & 0 \\ 0 & Z_{\a'} },
 \label{spinor:as}
\end{eqnarray}
with $C^{\dag}=C$ and $D^t=-D$, respectively. 
Here, 
\be
\Xi_{\a} = \mat{\xi_{\a}^{1} &~&~&~ \\ &~ \xi_{\a}^{2} &~&~ \\ 
                &~&~ \ddots &~ \\ &~&~&~ \xi_{\a}^{k} }, ~~~~~
Z_{\a'} = \mat{\zeta_{\a'}^{1} &~&~&~ \\ &~ \zeta_{\a'}^{2} &~&~ \\
                &~&~ \ddots & \\ &~&~&~ \zeta_{\a'}^{k} }
\ee
The spinorial indices $\a,\cdots$ imply that the matrices in eq. (\ref{spinor:ad}) 
 belong to the defining representation of the $usp(2k)$ Lie algebra 
 while the indices $\a',\cdots$ imply that the matrices in eq.(\ref{spinor:as}) 
 belong to the  antisymmetric representation. 
The fermionic part of the quadratic action then becomes 
\begin{align}
 S_f^{(2)}   &= -\frac{1}{g^2} \sum_{i,j} \left[ 
           	  \ol{O}^{ji} \G^{\mu} x_{\mu}^{ij} O^{ij} 
           	+ \ol{P^*}^{ij} \G^{\mu} \xt_{\mu}^{ij} P^{ij} 
           	+ \ol{C}^{ji} \G^{\mu} x_{\mu}^{ij} C^{ij} 
           	+ \ol{D^*}^{ij} \G^{\mu} \xt_{\mu}^{ij} D^{ij}
             \right] \cr 
             &~~~
             	-\frac{1}{g^2} \sum_{i,j} \left[
                  \ol{C}^{ji} \G^{n} x_{n}^{ij} O^{ij} 
                + \ol{O}^{ji} \G^{n} x_{n}^{ij} C^{ij}
             \right] 
             	-\frac{1}{g^2} \sum_{i,j} \left[
                  \ol{D^*}^{ij} \G^{n} x_{n}^{ij} P^{ij} 
                + \ol{P^*}^{ij} \G^{n} x_{n}^{ij} D^{ij}
             \right] \cr 
             &~~~
             	+ \frac{1}{g^2} \sum_{i,j} \left[
                  \ol{\xi}^{ij} \G^{\mu} O^{ij} M_{\mu}^{ij *} 
             	+ \ol{O}^{ji} \G^{\mu} \xi^{ij} M_{\mu}^{ij} 
             \right]
             	+ \frac{1}{g^2} \sum_{i,j} \left[
                  \ol{\zeta}^{ij} \G^{\mu} C^{ij} M_{\mu}^{ij*} 
             	+ \ol{C}^{ji} \G^{\mu} \zeta^{ij} M_{\mu}^{ij} 
             \right] \cr 
             &~~~
             	+ \frac{1}{g^2} \sum_{i,j} \left[
                  \ol{\xi}^{ij} \G^{n} C^{ij} A_{n}^{ij*} 
             	+ \ol{C}^{ji} \G^{n} \xi^{ij} A_{n}^{ij} 
             \right]
             	+ \frac{1}{g^2} \sum_{i,j} \left[
                  \ol{\zeta}^{ij} \G^{n} O^{ij} A_{n}^{ij*} 
             	+ \ol{O}^{ji} \G^{n} \zeta^{ij} A_{n}^{ij} 
             \right] \cr 
             &~~~
             	+\frac{1}{g^2} \sum_{i,j} \left[
                  \ol{\xit}^{ij} \G^{\mu} P^{ij} N_{\mu}^{ij*} 
             	+ \ol{P^*}^{ij} \G^{\mu} \xit^{ij} N_{\mu}^{ij} 
             \right]
             	+\frac{1}{g^2} \sum_{i,j} \left[
                  \ol{\zeta}^{ij} \G^{n} P^{ij} B_{n}^{ij*} 
             	+ \ol{P^*}^{ij} \G^{n} \zeta^{ij} B_{n}^{ij} 
             \right] \cr 
             &~~~
             	+\frac{1}{g^2} \sum_{i,j} \left[
                  \ol{\zeta}^{ij} \G^{\mu} D^{ij} N_{\mu}^{ij*} 
             	+ \ol{D^*}^{ij} \G^{\mu} \zeta^{ij} N_{\mu}^{ij} 
             \right]
             	+\frac{1}{g^2} \sum_{i,j} \left[
                  \ol{\xit}^{ij} \G^{n} D^{ij} B_{n}^{ij*} 
             	+ \ol{D^*}^{ij} \G^{n} \xit^{ij} B_{n}^{ij} 
             \right], \cr 
\label{quad:fermion}
\end{align}
where $\xi^{ij} = \xi^i -\xi^j$, $\xit^{ij} = \xi^i + \xi^j$ and $\z^{ij} = \z^i -\z^j$ 
 and the spinor indices are suppressed.
We have managed to represent $S_b + S_f$ as a sum of the following quadratic forms:  
\begin{align}
 S^{(2)} &=  \sum_{i,j} \frac{(x^{ij}_{M})^2}{g^2} 
 		\mat{M_{\mu}^{ij*} & A_{n}^{ij*}} 
                \mat{\eta^{\mu \nu} + \CT_{0(ij)}^{\mu\nu}& \CT_{1(ij)}^{\mu b} 
                \\ \CT_{1(ij)}^{n \mu} & \eta^{n b} + \CT_{0(ij)}^{n b}}
                \mat{M_{\nu}^{ij} \\ A_{b}^{ij}} \cr
         &~~~~~  \sum_{i,j} \frac{(\xt^{ij}_{M})^2}{g^2}  
 		\mat{N_{\mu}^{ij*} & B_{n}^{ij*}} 
                \mat{\eta^{\mu \nu} + \CTt_{0(ij)}^{\mu\nu} & \CTt_{1(ij)}^{\mu b} 
                \\ \CTt_{1(ij)}^{n \mu} & \eta^{n b} + \CTt_{0(ij)}^{n b}}
                \mat{N_{\nu}^{ij} \\ B_{b}^{ij}} \cr
         &~~~~~ -\frac{1}{g^2} \sum_{i,j} 
 		\left\{ \ol{O}^{ji} + \cdots \right\} \CS_0^{(ij)}
        	\left\{ {O}^{ij} + \cdots \right\} \cr 
 	 &~~~~~ -\frac{1}{g^2} \sum_{i,j} 
 		\left\{ \ol{C}^{ji} + \cdots \right\} 
        	\CS_1^{(ij)}
        	\left\{ {C}^{ij} + \cdots \right\} \cr 
         &~~~~~ -\frac{1}{g^2} \sum_{i,j} 
 		\left\{ \ol{P^{*}}^{ij} + \cdots \right\} \CSt_0^{(ij)}
        	\left\{ {P}^{ij} + \cdots \right\} \cr 
 	 &~~~~~ -\frac{1}{g^2} \sum_{i,j} 
 		\left\{ \ol{D^{*}}^{ij} + \cdots \right\} 
        	\CSt_1^{(ij)}
        	\left\{ {D}^{ij} + \cdots \right\}, 
\label{quad}
\end{align}
where
\begin{align}
 \CS_0^{(ij)} &\equiv (\G^{\mu} x_{\mu}^{ij}), ~~~~ 
 \CS_1^{(ij)} 
      \equiv \G^{\mu}x_{\mu}^{ij} \frac{(x_{M}^{ij})^2}{(x_{\mu}^{ij})^2}, ~~~~
\end{align}
and 
\begin{align}
 \CT_{0(ij)}^{MN} 
  &= \frac{1}{(x_P^{ij})^4} \bar{\xi}_{(+)}^{ij}  
     \mat{ {P}_{\ad} \G^M (x_{\mu}^{ij} \G^{\mu}) \G^N P_{\ad} ~ & ~ 
     	   {P}_{\ad} \G^M (x_{n}^{ij} \G^{n}) \G^N P_{\as} \\ 
           {P}_{\as} \G^M (x_{n}^{ij} \G^{n}) \G^N P_{\ad} ~ & ~ 
           {P}_{\as} \G^M (x_{\mu}^{ij} \G^{\mu}) \G^N P_{\as}}  
     \xi_{(+)}^{ij}, 
 \label{T_0(ij)} \\
 \CT_{1(ij)}^{MN} 
  &= \frac{1}{(x_P^{ij})^4} \bar{\xi}_{(+)}^{ij} 
     \mat{{P}_{\ad} \G^M (x_{n}^{ij} \G^{n}) \G^N P_{\ad} ~ & ~ 
          {P}_{\ad} \G^M (x_{\mu}^{ij} \G^{\mu}) \G^N P_{\as} \\ 
          {P}_{\as} \G^M (x_{\mu}^{ij} \G^{\mu}) \G^N P_{\ad} ~ & ~ 
          {P}_{\as} \G^M (x_{n}^{ij} \G^{n}) \G^N P_{\as}} 
     \xi_{(+)}^{ij}, 
 \label{T_1(ij)}
\end{align}
with $\xi_{(+)}^{ij} = ( \xi_{\a}^{ij} , \zeta_{\a'}^{ij} )^t$.   
Here we have introduced the projectors $P_{\ad}$  and $P_{\as}$, 
\be 
 P_{\ad} \xi^{ij}_{(+)} = \mat{\xi^{ij}_{\a} \\ 0}, ~~~ 
 P_{\as} \xi^{ij}_{(+)} = \mat{0 \\ \z^{ij}_{\a'}}.  
\ee
$\tilde{\CS}$'s and $\tilde{\CT}$'s are obtained from 
 $\CS$'s and $\CT$'s, replacing 
$x_{\mu}^{ij} = x_{\mu}^i - x_{\mu}^j$ and  $\xi_{\a}^{ij} 
              = \xi_{\a}^i - \xi_{\a}^j$ 
 by $\xt_{\mu}^{ij} = x_{\mu}^i + x_{\mu}^j$ 
 and $\xit_{\a}^{ij} = \xi_{\a}^i + \xi_{\a}^j$  respectively. 
We introduce $\xit_{(+)}^{ij} = ( \xit_{\a}^{ij} , \zeta_{\a'}^{ij} )^t$. 

We now sketch the derivation of (\ref{quad}), which is essentially ``completing squares". 
We will, in particular, show how $\CS_0$ and $\CS_1$ appear in the expression.
The fermionic shifts omitted as ``$\cdots$" in (\ref{quad}) are obtained 
 in the following procedure: for the simplicity 
let us  illustrate this in the quadratic terms consisting of $O^{ij}$ and 
 $C^{ij}$ in (\ref{quad:fermion}), 
\begin{align}
S^{(2)}_f[O,C;2] =  &-\frac{1}{g^2} \sum_{i,j} \left[ 
           	  \ol{O}^{ji} \G^{\mu} x_{\mu}^{ij} O^{ij} 
           	+ \ol{C}^{ji} \G^{\mu} x_{\mu}^{ij} C^{ij} 
             \right] 
             	-\frac{1}{g^2} \sum_{i,j} \left[
                  \ol{C}^{ji} \G^{n} x_{n}^{ij} O^{ij} 
                + \ol{O}^{ji} \G^{n} x_{n}^{ij} C^{ij}
             \right]. 
\label{s_f:O,C;2}
\end{align}
The quadratic terms consisting of $P^{ij}$ and $D^{ij}$ are similar to those of 
 $O^{ij}$ and $C^{ij}$. 
Eq. (\ref{s_f:O,C;2}) is rewritten as 
\begin{align}
S_f^{(2)}[O,C;2] = 
&-\frac{1}{g^2} \sum_{i,j} 
           	  (\ol{O}^{ji} + \ol{C}^{ji} \G^{n} x_{n}^{ij} (\G^{\mu} x_{\mu}^{ij})^{-1}) 
                  \G^{\mu} x_{\mu}^{ij} 
                  (O^{ij} + (\G^{\mu} x_{\mu}^{ij})^{-1} \G^{n} x_{n}^{ij} C^{ij})  \cr
             &~~~ 
                 - \frac{1}{g^2} \ol{C}^{ji} 
                 (\G^{\mu} x_{\mu}^{ij} - \G^{n} x_{n}^{ij}  
                     (\G^{\mu} x_{\mu}^{ij})^{-1}  \G^{n} x_{n}^{ij} )
                 C^{ij}. 
\label{OC}
\end{align}
Using 
\be
 (\CS_0^{(ij)})^{-1} = (\G^{\mu} x_{\mu}^{ij})^{-1} 
 = \frac{\G^{\mu}x_{\mu}^{ij}}{(x_{\mu}^{ij})^2}, 
\ee 
we obtain 
\begin{align}
\G^{\mu} x_{\mu}^{ij} - \G^{n} x_{n}^{ij}  
                     (\G^{\mu} x_{\mu}^{ij})^{-1}  \G^{n} x_{n}^{ij} 
                     = \G^{\mu}x_{\mu}^{ij}\frac{(x_{M}^{ij})^2}{(x_{\mu}^{ij})^2} 
                     = \CS_1^{(ij)}.
\end{align} 
When we include the linear terms in $O^{ij}$ and $C^{ij}$ in (\ref{quad:fermion}), 
 the fermions get shifted, but they have no effect on 
 $\CS_0^{(ij)}$ and $\CS_1^{(ij)}$.
$\CS_{0}^{(ij)}$  depends only on the adj. directions and 
 $\CS_{1}^{(ij)}$ is not seen in the calculation at the original IIB matrix model.  

Performing the integrations with respect to the fluctuations and 
 taking into account the ghost part which accompanies the gauge fixing (\ref{gauge fix}), 
 we obtain 
\begin{align}
e^{-S_{\eff}^{\text{1-loop}}[x,\eta]}  
 &= \int d\vt d\Pst db dc e^{-(S^{(2)} + S_{g.f.} + S_{\text{ghost}})} \cr
 &= \prod_{i,j(i\neq j)} {\det (\eta^{MN} + T^{MN}_{(ij)})^{-1}} 
 {\det (\eta^{MN} + T^{MN}_{(i\bar{j})})^{-1}} 
 \prod_{i} 
 {\det (\eta^{\mu\nu} + T^{\mu\nu}_{(i\bar{i})})^{-1} }.
\label{free energy}\cr
\end{align}
\be
T_{(ij)}^{MN} = \mat{\CT_{0(ij)}^{\mu \nu} & \CT_{1(ij)}^{\mu m} \\ 
                     \CT_{1(ij)}^{n \nu} & \CT_{0(ij)}^{nm}}
 = S_{(ij)}^{MN}  - U_{(ij)}^{MN}, 
\label{T_{ij}:USp}
\ee 
where 
\begin{align}
S_{(ij)}^{MN} &= \frac{1}{(x^{ij})^4} \bar{\xi}_{(+)}^{ij} 
 \left[ \G^{MPN} x_P^{ij} \right] \xi_{(+)}^{ij},  
\label{S_{ij}:IIB}
\end{align} 
and 
\be
 U_{(ij)}^{MN} = \mat{\CT_{1(ij)}^{\mu \nu} & \CT_{0(ij)}^{\mu m} \\ 
 		      \CT_{0(ij)}^{n\nu} & \CT_{1(ij)}^{nm}}. 
\ee 
Here the matrices and the determinants are with respect to the Lorentz indices. 
Eq.(\ref{S_{ij}:IIB}) is the matrix \cite{AIKKT} which have appeared 
 in the calculation at the IIB matrix model. 
We conclude that eq.(\ref{T_{ij}:USp}) is the first effect of matrix orientifolding.

In (\ref{free energy}), we have introduced the notation 
\be
 T_{(i\bar{j})} =\tilde{T}_{(ij)}, 
 \label{int:i-m}
\ee
We mean by this that $\bar{j}$ denotes the mirror image of ${j}$. 
It is understood that $T_{i\bar{j}}$ corresponds to 
 the interaction between  point $i$ and its mirror image ${\bar{j}}$. 
As taking mirror image implies $x^{i}_{\mu} \to -x^{i}_{\mu}$ 
 and $x^{i}_{n} \to x^{i}_{n}$, this interpretation is justified. 

On the other hand, since 
\be
T^{\mu\nu}_{(i\bar{i})} = \tilde{\CT}_{0(ii)}^{\mu \nu}, 
\label{T_{ii}:USp}
\ee 
where 
\begin{align}
 \CTt_{0(ii)}^{\mu\nu} 
  &= \frac{\tilde{x}_{\l}^{ii}}{(\xt_{\s}^{ii})^4}  
     \bar{\xit}_{(+)}^{ii} 
     \mat{P_{\ad} \G^{\mu} \G^{\l} \G^{\nu} P_{\ad} ~ & ~ 0 \\ 
     0 ~ & ~ P_{\as} \G^{\mu} \G^{\l} \G^{\nu} P_{\as}} 
     \xit_{(+)}^{ii} 
   = \frac{\tilde{x}_{\l}^{ii}}{(\xt_{\s}^{ii})^4}  
     \bar{\xit}^{ii}_{(+)} 
     \G^{\mu} \G^{\l} \G^{\nu} 
     \xit^{ii}_{(+)}, 
  \label{T_0(ii)}   
\end{align} 
and the third determinant in (\ref{free energy}) 
 $\det (\eta^{\mu\nu} + T^{\mu\nu}_{(i\bar{i})})^{-1}$ has no dependence 
 on the antisymmetric direction. 
Note that $\xit^{ii}_{(+)} = (\xit^{ii}_{\a},0)^t$. 

\begin{figure}[h]
\begin{center}
\includegraphics[scale=0.5]{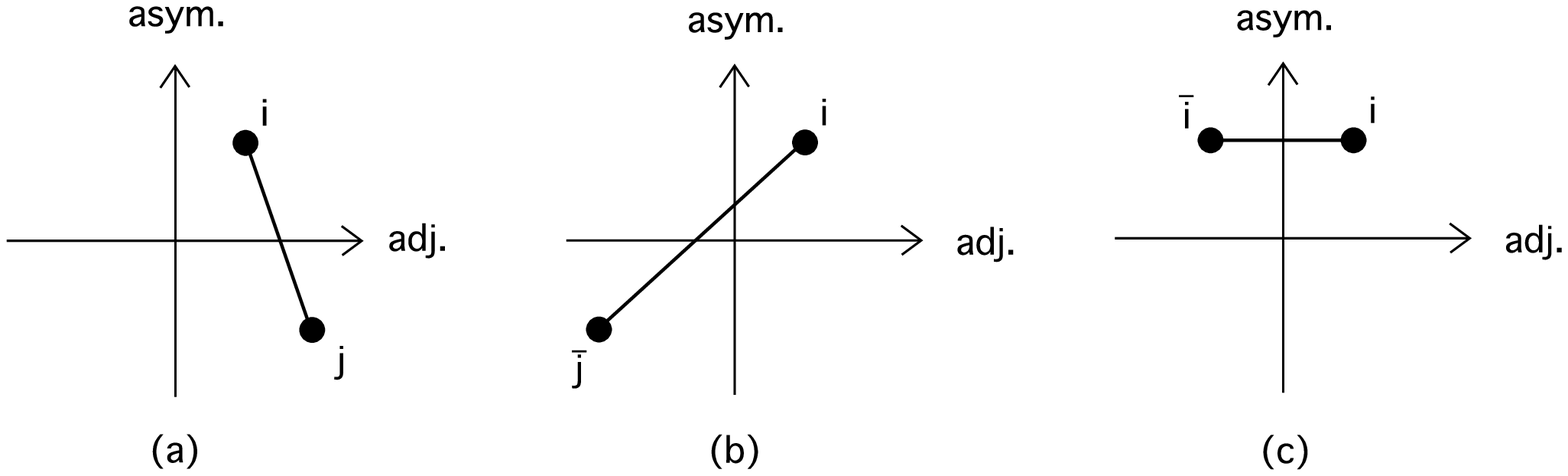} 
\caption{{\bf (a)}~~$T_{ij}$ ~~~ 
 {\bf (b)}~~$T_{i\bar{j}}=\tilde{T}_{ij}$ 
 ($x_{\mu}^{ij} \to \xt_{\mu}^{ij}$, $\xi_{\a}^{ij} \to \xit_{\a}^{ij}$) 
 {\bf (c)}~~$T_{i\bar{i}}=\tilde{T}_{ii}$ }
\label{T:fig}
\end{center}
\end{figure}

When we use the notation (\ref{int:i-m}), (\ref{free energy}) can be 
 collectively written as  
\be
 e^{-S_{\eff}^{\text{1-loop}}[x,\eta]} 
  = \prod_{i\neq j} {\det (\eta^{MN} + T^{MN}_{(ij)})^{-1}} 
    \prod_{i} {\det (\eta^{\mu\nu} + T^{\mu\nu}_{(i\bar{i})})^{-1} }. 
 \label{PF[x,y]}
\ee
From now, we will regard that 
 the product $\prod$ and the summation $\sum$ 
 are taken over all indices including these with bar. 
There are illustrated in Figure \ref{T:fig}. 
We conclude that this is the second effect of matrix orientifolding. 

We now turn to the issue of the $\xi$ integrations. 
We introduce graphical rules such that the outcome of the $\xi$ integrations
 is understood as a sum of all possible graphs generated.
To illustrate the situation, we begin with cases of small $k$. 

\begin{figure}[h]
\begin{center}
\includegraphics[scale=0.5]{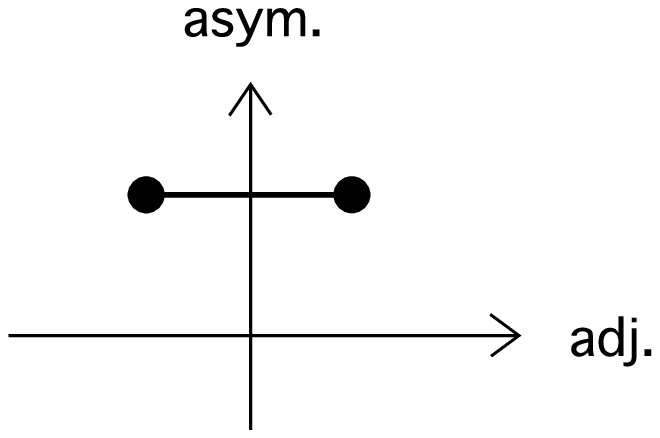} 
\caption{}
\label{usp(2):fig}
\end{center}
\end{figure}

\noindent
$\bullet USp(2)$ case.
  The diagonal matrices which have remained unintegrated are
\bea
  x_{\mu} = \mat{x_{\mu}^{1} & 0 \\ 0 & - x_{\mu}^{1}}, ~
  x_{m} = \mat{x_{m}^{1} & 0 \\ 0 &  x_{m}^{1}}, ~
  \eta_{\a} = \mat{\xi_{\a}^{1} & 0 \\ 0 & - \xi_{\a}^{1}}, ~
  \eta_{\a'} = \mat{\z_{\a'}^{1} & 0 \\ 0 &  \z_{\a'}^{1}}. 
\eea
 In eq. (\ref{f:2}), $S_f^{(2)}$, $\eta_{\alpha^{\prime}}$ lying 
  in the antisymmetric representation disappear as it is in the commutator. 
 As a result, the integrations reduce to those of the
   $D=6 ~ SU(2)$ matrix model \cite{AIKKT, ST}. 
Associated with $\eta_{\alpha}$ integrations, we draw a graph in Figure \ref{usp(2):fig} 
  which  consists of a single horizontal bond of multiplicity eight 
  (namely, a bond consisting of eight solid lines) connecting the point and its image.

\begin{figure}[h]
\begin{center}
\includegraphics[scale=0.5]{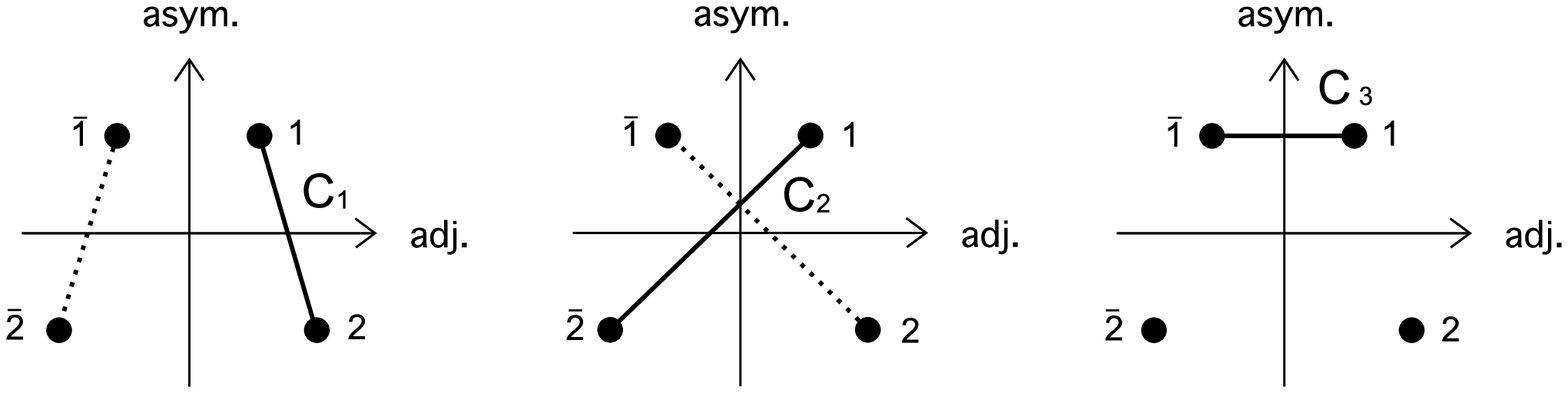} 
\caption{}
\label{usp(4)ad:fig}
\end{center}
\end{figure}

\noindent
$\bullet USp(4)$ case.
 The diagonal matrices are
\bea
  x_{\mu} = \mat{ x_{\mu}^{1} &&& \\ &x_{\mu}^{2}&& \\ 
                  && -x_{\mu}^{1} & \\ &&& -x_{\mu}^{2} },  &&
  x_{m} = \mat{ x_{m}^{1} &&& \\ &x_{m}^{2}&& \\ 
                  && x_{m}^{1} & \\ &&& x_{m}^{2} },  \cr
  \eta_{\a} = \mat{ \xi_{\a}^{1} &&& \\ &\xi_{\a}^{2}&& \\ 
                  && -\xi_{\a}^{1} & \\ &&& -\xi_{\a}^{2} },  &&
  \eta_{\a'} = \mat{ \z_{\a'}^{1} &&& \\ &\z_{\a'}^{2}&& \\ 
                  && \z_{\a'}^{1} & \\ &&& \z_{\a'}^{2} }. 
\eea
The overall $U(1)$ factors of the matrices in the antisymmetric representation decouple 
 and we can regard $\eta_{\alpha^{\prime}}$ effectively traceless: 
 $\zeta^2_{\alpha^{\prime}}= -\zeta^1_{\alpha^{\prime}}$. 
The integrations of our interest are
\beq
  \int \prod_{\a,\a'} d\xi_{\a}^{1} d\xi_{\a}^{2} d\z_{\a'}^{1} 
   e^{-S^{\text{1-loop}}_{\eff}[x,\eta]}. 
  \label{int:usp(4)}
\eeq
Let us first note  a single grassmann integration of $\xi_{\alpha}^{1}$ 
 which takes the form of
\beq
 \int d\xi_{\a}^{1} \left[ C_{0} + \sum_{\a} C_{1\a} (\xi_{\a}^{1} - \xi_{\a}^{2}) 
 		     + \sum_{\a} C_{2\a} (\xi_{\a}^{1} + \xi_{\a}^{2}) 
                     + \sum_{\a} C_{3\a} (\xi_{\a}^{1} + \xi_{\a}^{1}) \right]. 
\eeq
This integration is represented as a sum of the three graphs 
 each having a single solid line which are depicted in Figure \ref{usp(4)ad:fig}. 

\begin{figure}[h]
\begin{center}
\includegraphics[scale=0.55]{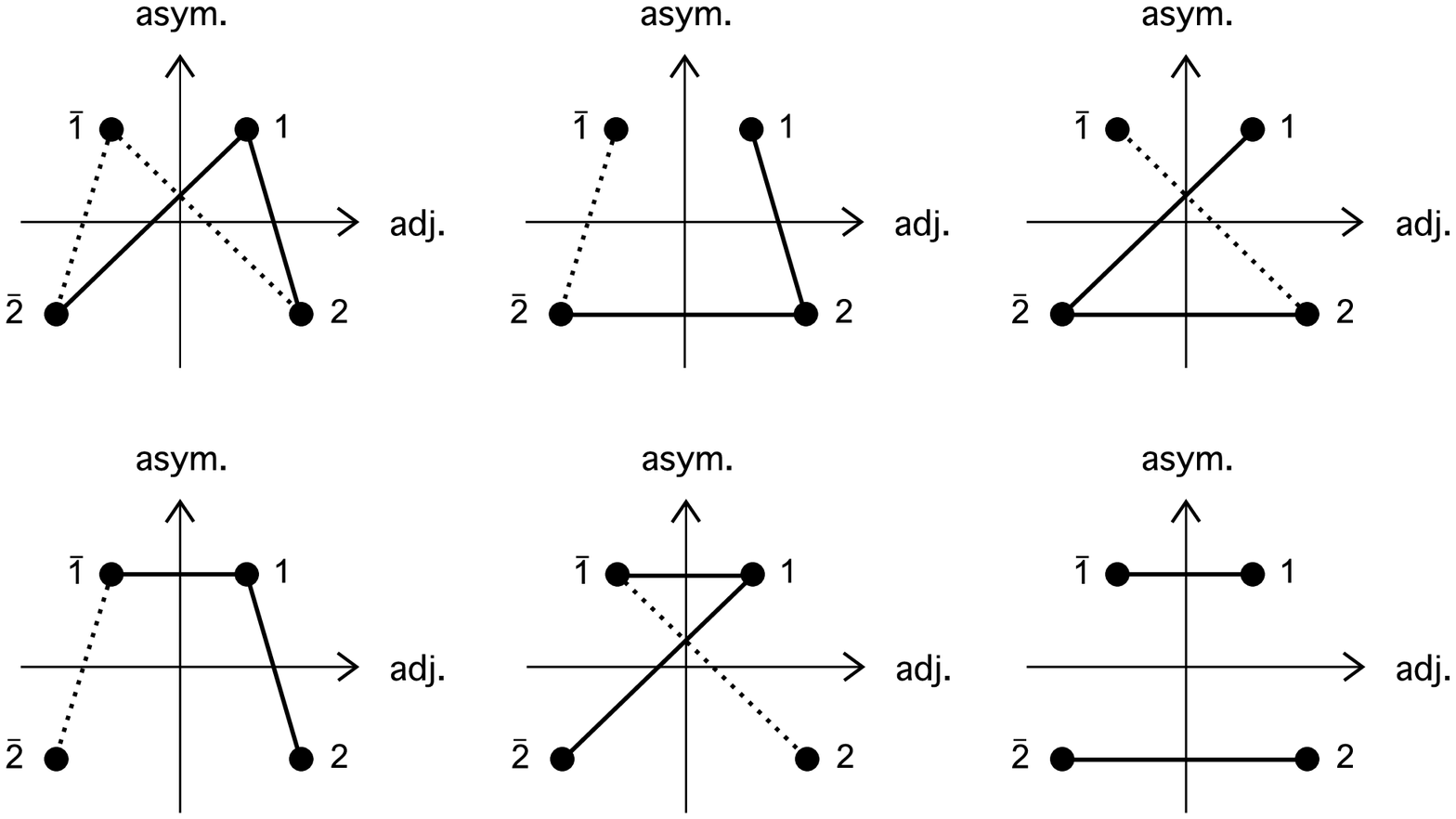} 
\caption{}
\label{usp(4)ad2:fig}
\end{center}
\end{figure}

Note that the image points are not independent dynamical degrees of freedom 
 but nonetheless required to represent our integrations. 
We have, therefore, drawn a dotted line to the image of each solid line except the case 
 that the solid line and its image are the same. 
Let us subsequently carry out the second  integration of $\xi_{\alpha}^{2}$. 
The result is expressed as a sum of the six graphs depicted in Figure \ref{usp(4)ad2:fig}. 

\begin{figure}[h]
\begin{center}
\includegraphics[scale=0.5]{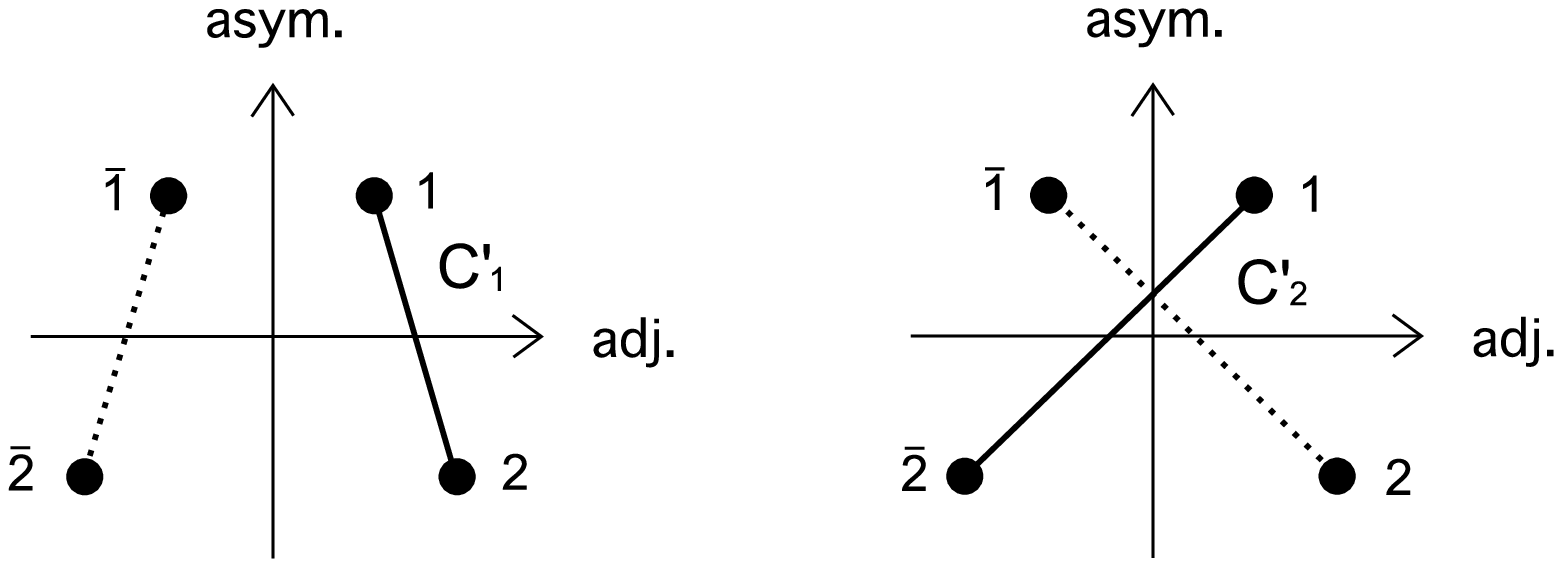} 
\caption{}
\label{usp(4)as:fig}
\end{center}
\end{figure}

\noindent
On the other hand, $\zeta^1_{\alpha^{\prime}}$ integration takes the form of 
\beq 
  \int d\z_{\a'}^{1} \left[ C_{0}' 
                      + \sum_{\a'} C_{1\a'}^{\prime} (\z_{\a'}^{1} - \z_{\a'}^{2}) 
  	 	      + \sum_{\a'} C_{2\a'}^{\prime} (\z_{\a'}^{1} - \z_{\a'}^{2}) 
                      \right], 
\eeq
and  is represented as a sum of the two graphs depicted in Figure \ref{usp(4)as:fig}.
  The integrations eq. (\ref{int:usp(4)}) are  understood as follows: 
  pick one graph from Figure \ref{usp(4)ad2:fig} and superpose 
  on top of it one graph from Figure \ref{usp(4)as:fig} for a given spinorial index. 
We repeat this procedure eight times to exhaust the spinorial indices and 
superpose the eight sheets  on top of each other to generate all graphs.

\noindent
Knowing these, we can proceed easily to the graphical rules to the general cases. 
In the case of $USp(2k)$, we draw all possible graphs consisting of $k$ solid lines 
 without loop and their mirror images drawn by dotted lines 
 for a given spinorial index in the adj representation.  
For a given spinorial index in the antisymmetric representation, 
 we do the same except that there are only $k-1$ solid lines  and that  
 we do not draw a solid line between a point and its image. 
We repeat this process eight times to exhaust the spinorial indices and 
 superpose  on top of each other to generate all possible graphs. 
There are, of course,  symmetries associated with the $6+4$ dimensional 
 Lorentz indices seen in eq. (\ref{PF[x,y]})  which, upon expansion, 
 result in structure of multiplicities of a bond, namely the one between 
 two points consisting a multiple of solid lines in a given final graph. 
While this is of some interest, it is outside the issue we raise in this paper 
 and will not be discussed here.

\section{Spacetime directional asymmetry of force exerting among diagonal elements}

We move on to discuss the qualitative features of the force among the diagonal elements 
 induced by the integration of off-diagonal elements. 
For that purpose, we first discuss cases where the size of the matrices is small.

\subsection{USp(2) case}

This is the simplest case. 
There is one spacetime point and its image  as illustrated 
 in Figure. \ref{lower_rank:fig} (1).
The matrices are 
\begin{eqnarray}
 v_{\mu} = \mat{x_{\mu} & m_{\mu} \\ m_{\mu}^{*} & -x_{\mu}}, &~~&
 v_{m} = \mat{x_{m} & 0 \\ 0 & x_{m}}, \\ 
 \Psi_{\a} = \mat{\xi_{\a} & p_{\a} \\ p_{\a}^{*} & -\xi_{\a}}, &~~&
 \Psi_{\a'} = \mat{\zeta_{\a'} & 0 \\ 0 & \zeta_{\a'}}. 
\end{eqnarray}
We represent $x_{\mu}$, $m_{\mu}$ and $x_m$ by six and four dimensional vectors 
 $\vec{x}$, $\vec{m}$ and $\vec{\vec{x}}$ respectively. 
By fixing the spacetime symmetry $SO(6)\times SO(4)$, 
\begin{align}
\vec{x} \to \mat{\frac{R_6}{2} \\ {\bf 0}_5}, ~~~
\vec{{m}} \to \mat{c_0 \\ c_1 \\ c_2\\ {\bf 0}_3}, ~~~
\vec{\vec{x}} \to \mat{\frac{R_4}{2} \\ {\bf 0}_3}. 
\end{align}
The $v_m$ lie in the overall $U(1)$ and decouples from the action.
Hence there is no dependence on $R_4$ in the effective action $S_{\eff}(R_6,R_4)$. 
The force by construction exerts in the adj. directions only. 
In fact, 
\begin{align}
 \int R_6^5 dR_6  \int R_4^3 dR_4 e^{-S_{\eff}(R_6,R_4)} 
 	 & \equiv \int dv_{\mu} d\Psi_{\a} e^{-S} \cr 
   	& \sim  
          \int dv_{\mu} \left( \det(\G^{\mu} \ad(v_{\mu}) )^{\frac{1}{2}} \right) 
          e^{ \frac{1}{4g^2} \tr[v_{\mu}, v_{\nu}] }.
\end{align}
The calculation of the effective action is the same as that 
 in the $D=6$ $SU(2)$ matrix model. 
The long and short distance behavior are already given. (\cite{AIKKT}. See also \cite{ST}.)
\begin{eqnarray}
 S_{\eff} ~~ \sim ~~
 &12 \log R_6, ~~~\text{for}~~ & R_6 \gg g^2, \cr
 &-4 \log R_6, ~~~\text{for}~~ & R_6 \ll g^2. 
 \label{usp(2):int} 
\end{eqnarray}
This represents a two-body force between the point and its mirror image which 
 is attractive in the long distance and repulsive in the short distance. 
There is an obvious directional asymmetry in the original ten dimensional sense.

\subsection{USp(4) case}
The matrices are 
\begin{eqnarray}
 v_{\mu} = \mat{x^1_{\mu}   & m_{\mu}   & n^1_{\mu} & n_{\mu}   \\ 
 		m_{\mu}^{*} & x^2_{\mu} & n_{\mu}   & n^2_{\mu} \\ 
                n^{1*}_{\mu} & n_{\mu}^{*} & -x^1_{\mu} & m_{\mu}^{*} \\
                n_{\mu}^{*}  & n^{2*}_{\mu} & m_{\mu} & -x^2_{\mu}}
 , &&~~~
 v_{m} = \mat{x^1_{m}   & a_{m}   & 0 & b_{m}   \\ 
 		a_{m}^{*} & x^2_{m} & -b_{m}   & 0 \\ 
                0 & -b_{m}^{*} & x^1_{m} & a_{m}^{*} \\
                b_{m}^{*}  & 0 & a_{m} & x^2_{m}}, \\ 
 \Psi_{\a} = \mat{\xi^1_{\a}   & o_{\a}   & p^1_{\a} & p_{\a}   \\ 
 		  o_{\a}^{*} & \xi^2_{\a} & p_{\a}   & p^2_{\a} \\ 
                  p^{1*}_{\a} & p_{\a}^{*} & -\xi^1_{\a} & -o_{\a}^{*} \\
                  p_{\a}^{*}  & p^{2*}_{\a} & -o_{\a} & -\xi^2_{\a}}
 , &&~~~
 \Psi_{\a'} = \mat{\z^1_{\a'}   & c_{\a'}   & 0 & d_{\a'}   \\ 
 		   c_{\a'}^{*} & \z^2_{\a'} & -d_{\a'}   & 0 \\ 
                   0 & -d_{\a'}^{*} & \z^1_{\a'} & c_{\a'}^{*} \\
                   d_{\a'}^{*}  & 0 & c_{\a'} & \z^2_{\a'}}. 
\end{eqnarray}
The exact multiple integrations to obtain 
 $S_{\eff}(R_6^{(ij)},R_4^{(ij)};i,j=1,\bar{1},2,\bar{2})$ are difficult to perform 
 and we will not attempt it here. 
Leaving aside these many body effects among the spacetime points, we content ourselves 
 to see what kind of two-body forces are exerting by ignoring some of the integrations. 
 
If we keep $\{m_{\mu}, a_{n}, o_{\a}, c_{\a'} \}$ and ignore the remaining off-diagonals, 
 we obtain the two-body force between 1 and 2. 
The calculation is the same as that of $D=10$ $SU(2)$ matrix model: 
\begin{eqnarray}
 24 \log \sqrt{R_6^2 + R_4^2} && \text{at long distances}, \cr 
 -8 \log \sqrt{R_6^2 + R_4^2} && \text{at short distances}. 
 \label{S_eff:usp(4)}
\end{eqnarray}
The same is true if we keep $\{n_{\mu}, b_{n}, p_{\a}, d_{\a'} \}$, ignoring the remaining 
 off-diagonals, and therefore exhibiting the two-body force between 1 and $\bar{2}$ 
 of the same kind as above. 

The situation changes when we consider the two-body force between 
 1 and $\bar{1}$ (or 2 and $\bar{2}$). 
The diagonal elements of the off-diagonal block of the matrix belonging 
 to the antisymmetric representation are zero. 
The force of this type is generated by the integrations of $\{n^1_{\mu}, p^1_{\a}\}$ 
 (or $\{n^2_{\mu}, p^2_{\a}\}$). 
The calculation therefore becomes the same as the $USp(2)$ case 
 and we obtain (\ref{usp(2):int}).

\begin{figure}[h]
\begin{center}
\includegraphics[scale=0.6]{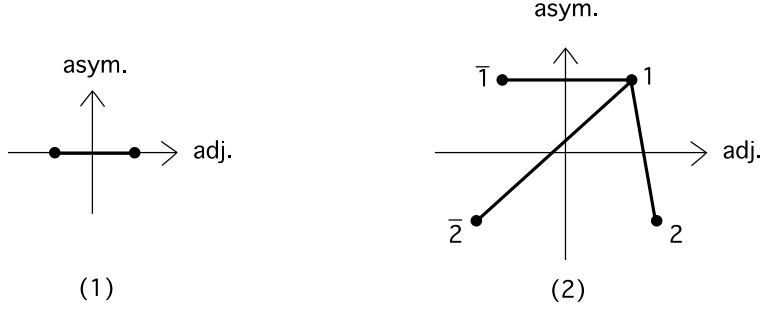} 
\caption{(1) USp(2)  (2) USp(4)}
\label{lower_rank:fig}
\end{center}
\end{figure}

We illustrate these three kinds of two-body interactions acting on point 1 by bonds 
 in Figure. \ref{lower_rank:fig} (2). 
At long distances, point 1 gets attracted not only by point 2, but also 
 vertically toward the four dimensional plane spanned by the directions lying 
 in the antisymmetric representation.

\subsection{Induced forces derived from the one-loop determinant}

Having these aspects of induced forces in mind, we come back to section three. 
In particular, let us examine eqs.(\ref{free energy}), (\ref{T_{ij}:USp}), (\ref{T_{ii}:USp}), 
 (\ref{T_0(ii)}). 
Suppose that we pick up $\det(\eta^{MN} + T^{MN}_{(ij)})$ upon integrations 
 over $d\xi^i$ and $d\z^i$.
Then ignoring the spinorial structure, and saturating the grassmann integrations, 
we see that the power behavior is 
\be
 \left( \frac{x^{ij}}{(x^{ij})^4} \right)^8 \sim \frac{1}{(R_{10})^{24}}. 
\ee
Here we have denoted $(x^{ij})^2 = x^{ijM}x^{ij}_M \equiv (R_6)^2 +(R_4)^2 = (R_{10})^2$. 
This is in accordance with eq.(\ref{S_eff:usp(4)}). 
The same is true when we pick up $\det(\eta^{MN} + T^{MN}_{(i\bar{j})})$. 
The power behavior appearing upon integrations is 
\be
 \left( \frac{x^{i\bar{j}}}{(x^{i\bar{j}})^4} \right)^8 \sim \frac{1}{(\tilde{R}_{10})^{24}}. 
\ee
Here  $(\tilde{R}_{10})^2 = (\tilde{R}_6)^2 + (R_4)^2 = x^{i\bar{j}M} x^{i\bar{j}}_M$.
On the other hand, when we pick up $\det(\eta^{\mu\nu} + T^{\mu\nu}_{(i\bar{i})})$, 
 we see that the modes $\z_i$ in the antisymmetric representation decouple and the 
 power behavior is 
\be
 \left( \frac{x^{i\bar{i}}}{(x^{i\bar{i}}_{\mu})^4} \right)^4 
 \sim \frac{1}{\tilde{R}_{6}^{12}}, 
\ee
which is in accordance with eq.(\ref{usp(2):int}). 
We have thus reached a picture consistent with that of section three.


{\bf{Acknowledgements}}\\
We thank Hajime Aoki, Hironobu Kihara, Kazutoshi Ohta and Asato Tsuchiya 
 for useful discussions.
This work is supported in part by the Grant-in-Aid for Scientific Research (2054278) 
from the Ministry of Education, Science and Culture, Japan.

\vspace{3mm}



\newpage

\begin{thebibliography}{99} 

\bibitem{BFSS}
T. Banks, W. Fishler, S. H. Schenker and L. Susskind, 
``M theory as a matrix model: A conjecture" 
Phys. Rev. D {\bf 55}: 5112, (1997) [hep-th/9610043]. 

\bibitem{IKKT}
N. Ishibashi, H. Kawai, Y. Kitazawa, A. Tsuchiya,
``A Large N reduced model as superstring" 
Nucl. Phys. B {\bf 498}: 467, (1997) [hep-th/9612115].

\bibitem{DVV}
R. Dijkgraaf, E. P. Verlinde and H. L. Verlinde, 
``Matrix string theory" 
Nucl. Phys. B {\bf 500}: 43-61 (1997) [hep-th/9703030]. 

\bibitem{IToku}
H. Itoyama and A. Tokura, 
``USp(2k) matrix model: F theory connection" 
Prog. Theor. Phys. {\bf 99}, 129 (1998) [hep-th/9708123]; 
H. Itoyama and A. Tokura, 
``USp(2k) matrix model: Nonperturbative approach to orientifolds", 
Phys. Rev. D {\bf 58}, 026002 (1998) [hep-th/9801084]. 

\bibitem{ITsu}
H. Itoyama and A. Tsuchiya, 
``USp(2k) matrix model" 
Prog. Theor. Phys. Suppl. 134, 18 (1999) [hep-th/9904018]; 
``USp(2k) matrix model: Schwinger-Dyson equations and closed open string interactions" 
Prog. Theor. Phys. {\bf 101}: 1371-1390 (1999) [hep-th/9812177]. 

\bibitem{DF}
U. H. Danielsson and G. Ferretti, 
``The Heterotic Life of the D Particle" 
Int. J. Mod. Phys. A {\bf 12}: 4581-4596 (1997) [hep-th/9610082]; 
L. Motl
``Quaternions and M(atrix) theory in spaces with boundaries" 
[hep-th/9612198];
N. Kim and S. J. Rey, 
``M(atrix) Theory on an Orbifold and Twisted Membrane" 
Nucl. Phys. B {\bf 504}: 189-213, (1997) [hep-th/9701139]. 

\bibitem{KSlv}
S. Kachru and E. Silverstein, 
``On Gauge Bosons in the Matrix Model Approach to M Theory" 
Phys. Lett. B {\bf 396}: 70-76, (1997) [hep-th/9612162]; 
D. A. Lowe, 
``Heterotic Matrix String Theory" 
Phys. lett. B {\bf 403}: 243-249, (1997) [hep-th/9704041]; 
T. Banks, N. Seiberg and E. Silverstein, 
``Zero and One-dimensional Probes with N=8 Supersymmetry" 
Phys. Lett. B {\bf 401}: 30-37, (1997) [hep-th/9703052]. 

\bibitem{Tay}
W. Taylor
``M(atrix) theory: matrix quantum mechanics as a fundamental theory" 
Rev. Mod. Phys. {\bf73}: 419 (2001) [hep-th/0101126].

\bibitem{IY}
H. Itoyama and R. Yoshioka, 
``Matrix orientifolding and models with four or eight supercharges"
Phys. Rev. D {\bf 72}, 126005 (2005) [hep-th/0509146]. 

\bibitem{AIS}
H. Aoki, S. Iso, and T. Suyama
``Orbifold Matrix Model"
Nucl. Phys. B {\bf 634}: 71-89 (2002) [hep-th/0203277];
A. Miyake
``Supersymmetric Matrix Model on Z-Orbifold"
Int. J. Mod. Phys. A {\bf 19}: 1983-1912 (2004) [hep-th/0305106]. 

\bibitem{Un}  
M. \"Unsal   
``Regularization of Non-commutative SYM by Orbifolds 
 with Discrete Torsion and SL(2,Z) Duality" 
J. High Energy Phys. {\bf 12}: 033 (2005) [hep-th/0409106]; 
D. B. Kaplan and M. \"Unsal 
``A Euclidean Lattice Construction of
 Supersymmetric Yang-Mills Theories with Sixteen Supercharges"  
J. High Energy Phys. {\bf 09}: 042 (2005) [hep-lat/0503039]; 
M. \"Unsal, 
``Twisted supersymmetric gauge theories and orbifold lattices" 
J. High Energy Phys. {\bf 10}: 089 (2006) [hep-th/0603046]. 

\bibitem{EK}
T. Eguchi and H. Kawai 
``Reduction of Dynamical Degrees of Freedom in the Large N Gauge Theory" 
Phys. Rev. Lett. {\bf 48}:1063 (1982). 

\bibitem{MNS}
G. W. Moore, N. Nekrasov and S. Shatashvili
``D particle bound states and generalized instantons"
Commun. Math. Phys. {\bf 209}: 77-95 (2000) [hep-th/9803265]. 

\bibitem{SS}
S. Sethi and M. Stern 
``D-brane bound states reduxh
Commun. Math. Phys. {\bf 194}: 675 (1998) [hep-th/9705046].

\bibitem{PR}
M. Porrati and A. Rozenberg  
``Bound states at threshold in supersymmetric quantum mechanicsh
Nucl. Phys. B {\bf 515}, 184 (1998) [hep-th/9708119].

\bibitem{GG}
M. B. Green and M. Gutperle 
``D-particle bound states and the D-instanton measureh 
J. High Energy Phys, {\bf 01}: 005 (1998) [hep-th/9711107].

\bibitem{Stau}
M. Staudacher
``Bulk Witten indices and the number of normalizable ground states in supersymmetric 
quantum mechanics of orthogonal, symplectic and exceptional groups" 
Phys. Lett. B {\bf 488}: 194-198 (2000) [hep-th/0006234]. 

\bibitem{Pes}
V. Pestun, 
``N=4 SYM matrix integrals for almost all simple gauge groups (except E(7) and E(8))" 
J. High Energy Phys. {\bf 09}: 012 (2002) [hep-th/0206069]. 

\bibitem{IKY}
H. Itoyama, H. Kihara and R. Yoshioka, 
``Partition Functions of Reduced Matrix Models with Classical Gauge Groups" 
Nucl. Phys. B {\bf 762}: 285 (2007) [hep-th/0609063]. 

\bibitem{Kih}
H. Kihara 
``Grand Partition Functions of Little Matrix Models with ABCD" 
arXiv:0803.3984 [hep-th].  

\bibitem{AIKKT}
H. Aoki, S. Iso, H. Kawai, Y. Kitazawa and T. Tada, 
``Space-Time Structure from IIB Matrix Model", 
Prog. Theor. Phys. {\bf 99}: 713, (1998) [hep-th/9802085].

\bibitem{CIK2}
B. Chen, H. Itoyama and H. Kihara, 
``Nonabelian monopoles from matrices: Seeds of the spacetime structure", 
Nucl. Phys. B {\bf 577}, 23 (2000) [hep-th/9909075].

\bibitem{IM}
H. Itoyama and T. Matsuo, 
``Berry's connection and USp(2k) matrix model,h
Phys. Lett. B {\bf 439}, 46 (1998) [hep-th/9806139];
B. Chen, H. Itoyama and H. Kihara, 
``Nonabelian Berry phase, Yang-Mills instanton and USp(2k) matrix model",
Mod. Phys. Lett. A {\bf 14}, 869 (1999) [hep-th/9810237].

\bibitem{GT}
G. W. Gibbons and P. K. Townsend, 
``Self-gravitating Yang Monopoles in all Dimensions" 
Class. Quant. Grav. {\bf 23}: 4873-4886 (2006) [hep-th/0604024]. 

\bibitem{PST}
C. Pedder, J. Sonner and D. Tong
``The Geometric Phase and Gravitational Precession of D-Branes" 
Phys. Rev. D {\bf 76}: 126014 (2007) arXiv:0709.2136 [hep-th]; 
``The Berry Phase of D0-Branes"
J. High Energy Phys. {\bf 03}: 065 (2008) arXiv:0801.1813 [hep-th]. 

\bibitem{ST}
T. Suyama and A. Tsuchiya, 
``Exact results in N(c) = 2 IIB matrix model", 
Prog. Theor. Phys. {\bf 99}: 321-325, (1998) [hep-th/9711073].  





\end{thebibliography}
\end{document}